\documentclass[
	a4paper, 
	10pt, 
	twoside, 
]{LTJournalArticle}

\usepackage{amsmath}
\usepackage{amssymb}
\usepackage{mathrsfs}
\usepackage{xcolor} 
\usepackage[normalem]{ulem} 

\usepackage{mathtools}
\DeclareMathOperator*{\argmin}{arg\,min}

\runninghead{} 

\footertext{} 

\title{Physics-Informed Neural Networks to Infer the Perpendicular Energy Conductivity in the Scrape-Off Layer of Stellarator Devices} 

\author{%
	J. Gallego\textsuperscript{1}, 
    P. Protopapas\textsuperscript{2}, 
    A. Bustos\textsuperscript{1}, 
    A. Alonso\textsuperscript{3}, 
    S. Barquero\textsuperscript{3}, 
    A. Baciero\textsuperscript{3}, 
    I. Rivera\textsuperscript{3},\\ 
    J. A. Moríñigo\textsuperscript{1},
    R. Mayo-García\textsuperscript{1}, 
    and the TJ-II team.
}

\date{\footnotesize\textsuperscript{\textbf{1}}Departamento de Tecnología, CIEMAT, Spain.\\
\footnotesize\textsuperscript{\textbf{2}}Harvard John A. Paulson School of Engineering and Applied Sciences, USA.\\
\footnotesize\textsuperscript{\textbf{3}}Laboratorio Nacional de Fusión, CIEMAT, Spain.}

\renewcommand{\maketitlehookd}{%
	\begin{abstract}
In this work, we develop an inverse Physics-Informed Neural Network (PINN) framework to infer the dependence of the scrape-off layer (SOL) perpendicular heat conductivity on plasma density and temperature, $\kappa_\perp(n,T)$. The method combines radial profile measurements of electron density and temperature with the residual of a reduced one-dimensional SOL transport equation, so that the inferred conductivity is constrained by both the measurements and the underlying transport model. Three neural networks are trained simultaneously: two reconstruct the temperature and density profiles as functions of the radial coordinate and transported power, while a third represents the effective conductivity as a function of the local density and temperature. The framework is first validated using synthetic data generated from a prescribed conductivity function, allowing the inferred $\kappa_\perp(n,T)$ to be compared directly with the ground truth. The model recovers the imposed functional dependence with errors below $10~\%$ in the data-constrained region. Bootstrap resampling is shown to provide a practical indicator of prediction reliability and consistency. A scan in the number of plasma profiles used for training and the number of radial measurement positions per profile identifies a practical trade-off between reconstruction accuracy and data availability. Finally, the method is applied to an experimental dataset from the TJ-II stellarator obtained with the helium-beam diagnostic. This exploratory application provides an initial estimate of the effective SOL conductivity and illustrates the potential of inverse PINNs for extracting transport information from plasma edge measurements.
         
    \end{abstract}
}

\newcommand{\mrm}[1]{\mathrm{#1}}
\begin{document}

\maketitle 


\section{Introduction}

Neural networks are powerful tools for processing and generalizing large amounts of data, with successful applications in fields such as computer vision \cite{traore2018deep}, healthcare \cite{singh2022deep}, and civil engineering \cite{fan2022machine}. However, purely data-driven approaches rely entirely on the available data to contain the information needed to describe the system. As a result, their performance can deteriorate when the data are scarce, sparse, or not sufficiently representative of the underlying behaviour \cite{bansal2022systematic}.\\

Physics-Informed Neural Networks (PINNs) \cite{raissi2019physics} address this limitation by constraining the neural-network solution with the physical laws governing the system. In the PINN framework, these laws are embedded in the loss function as additional terms, so that the network is trained to satisfy both the available data and the prescribed physics. Since the original PINN formulation, the field has grown rapidly \cite{lawal2022physics,martinez2025physics}, reflecting the increasing interest of the scientific community in this approach.\\

When the governing physics is expressed as a differential equation, or as a system of differential equations, a PINN can be interpreted as a neural-network-based differential equation solver. In this setting, the network approximates the solution, while the initial and boundary conditions are imposed through the training dataset. PINNs have been applied as differential equation solvers in a wide range of problems, including fluid dynamics \cite{botarelli2025using}, heat transfer in multiphase flows \cite{jalili2024physics}, wave physics \cite{pellegrin2022transfer} and magnetohydrodynamic equations \cite{van1995neural,jaillon2026physics}.\\

Beyond forward problems, PINNs can also be formulated to solve inverse Partial Differential Equations (PDE) problems. In this case, the objective is to infer unknown quantities in the governing equation from sparse measurements of the solution, as also discussed in \cite{raissi2019physics}. In the simplest formulation, the unknown parameter is treated as an additional trainable variable and optimized so that the predicted solution remains consistent with both the measurements and the physical model. This strategy has been used, for example, to infer unknown parameters in groundwater flow \cite{depina2022application} and biomechanical systems \cite{li2021physics}. A further extension consists of representing the unknown coefficient itself as the output of a neural network, allowing it to depend on selected independent variables. The inverse problem then becomes the identification of a functional relation rather than a scalar parameter, as studied in \cite{tartakovsky2020physics}.\\

Since the original formulation of PINNs, numerous methodological developments have been introduced to improve their capabilities and extend their range of applications. Flamant \textit{et al}. \cite{flamant2020solving} introduced the concept of solution bundles, in which relevant parameters of the governing differential equations are included as inputs to the neural network, enabling a single model to represent a family of solutions rather than a single realization. This approach has subsequently been applied in several cosmological studies \cite{gomez2025evolution,chantada2023cosmology,chantada2024faster}. Further developments have addressed the difficulties associated with stiff differential equations. Tarancón-Álvarez \textit{et al}. \cite{tarancon2025efficient} proposed multi-head training combined with unimodular regularization to improve the treatment of stiff systems and to generalize the network over a latent space of solutions, while Yepes \textit{et al}. \cite{yepes2026gradient} investigated the sensitivity of PINN performance to the choice of initial-condition embeddings in stiff problems. Advances in alternative training techniques have also been proposed; for example, Protopapas \textit{et al}. \cite{protopapas2026variational} introduced variational boosting to enable second-order optimization strategies. In parallel, increasing attention has been devoted to the theoretical characterization of PINN accuracy. Recent work has derived error bounds for specific classes of differential equations \cite{chantada2024exact}, providing a more rigorous framework for quantifying the reliability of PINN solutions and improving their physical interpretability.\\

Building on these foundations, this work explores the application of a PINN framework to the modelling and interpretation of the physics parameters observed in the periphery of magnetically confined fusion plasmas. Nuclear fusion is a promising energy source, owing to its abundant fuel and lack of green house gas emissions. However, the practical production of fusion energy remains to be demonstrated. One of the central challenges is ensuring plasma heat and particle exhaust in the plasma-facing components that is compatible with stable, high-performance operation while maintaining the resulting heat loads within the thermal limits of the materials. For a given amount of transported power, this heat flux depends on the width $\lambda_q$ of the Scrape-Off Layer (SOL) transport channel, which is set by the balance between perpendicular and parallel transport. Here, perpendicular and parallel refer to the direction relative to the magnetic field. Higher perpendicular transport leads to a wider transport channel and therefore a lower peak heat load on the target. Determining the perpendicular and parallel heat conductivities, together with their parametric dependencies, is therefore essential for accurately predicting the heat loads the target must withstand and its wet area.\\

The parallel heat conductivity $\kappa_\parallel$ is well described by the Spitzer--Härm conductivity, which can be derived from classical collisional transport theory \cite{PhysRev.89.977}. In contrast, the perpendicular conductivity $\kappa_\perp$ is largely governed by turbulent transport and is therefore difficult to derive from first principles. One should also consider that the two main types of magnetic confinement fusion devices, tokamaks and stellarators, differ in their confinement and stability properties, and therefore perpendicular SOL transport cannot necessarily be assumed to follow the same behaviour in both designs. Previous studies have examined different aspects of perpendicular SOL transport mainly in tokamak devices. For example, Carralero \textit{et al}. \cite{carralero2018role} analysed the role of filaments in perpendicular SOL transport, while Itoh \textit{et al}. \cite{itoh1994scaling} studied the consistency and implications of applying $\kappa_\perp$ scaling laws developed for the confined plasma region to the SOL. Eich \textit{et al}. \cite{eich2013scaling} found a multi-machine scaling law to directly predict $\lambda_q$ in tokamaks. Nevertheless, robust scalings for $\kappa_\perp$ and, consequently, for $\lambda_q$ in the SOL of stellarator devices remain open problems.\\

The objective of this work is therefore to advance the study of $\kappa_\perp$ in stellarator devices. Namely, we present a new methodology to infer the functional dependence of the perpendicular heat conductivity on plasma density $n$ and temperature $T$ in the SOL, $\kappa_\perp(n,T)$, using sparse measurements of the plasma profiles. To this end, we use an inverse PINN framework based on the minimization of the residual of a reduced 1D transport model, which we implement in Python using NeuroDiffEq \cite{chen2020neurodiffeq,liu2025recent}.\\

The paper is organized as follows. Section~\ref{sec:background} introduces the physical and computational background of the work. First, the heat transport physics of the SOL and existing scalings for $\kappa_\perp$ are reviewed. The section then defines formally the inverse PINN formulation for parameter-identification problems. Section~\ref{sec:PINN_model} presents the PINN model, including the specific PINN structure adopted to solve the problem and the optimization process. Section~\ref{sec:synthetic_validation} validates the framework using synthetic data generated from numerical solutions of the governing equation. The validation assesses the robustness of the method through scans in the number of plasma profiles and measurement positions. Section~\ref{sec:exp} applies the methodology to a small experimental dataset from the TJ-II stellarator, describing the experimental setup, the measured data, and the inferred transport results. Finally, section~\ref{sec:conclusions} summarizes the main conclusions and outlines directions for future work.\\

\section{Background}\label{sec:background}
This section provides an overview of the physical and computational background of the work. First, the heat-transport physics of the SOL are introduced, including the reduced transport model used in this study and the main $\kappa_\perp$ scalings reported in the literature. The inverse PINN formulation is then presented, beginning with the inference of a constant coefficient and subsequently extending the formulation to the inference of a functional dependence.

\subsection{Heat transport physics in the SOL}\label{subsec:sol_heat_transport}

\subsubsection{Heat transport model}\label{subsubsec:heat_transport_model}

Within the two-fluid framework, the SOL energy balance for ions and electrons can be written as \cite{feng2022review}
\begin{align}
\label{eq:two_fluid1}
\nabla \cdot \left(
\frac{5}{2}nu_\parallel T_\mrm{i} \mathbf{b}
- \kappa_{\parallel \mrm{i}} \nabla_{\parallel} T_\mrm{i}
- \kappa_{\perp \mrm{i}} \nabla_{\perp} T_\mrm{i}
\right)
&= S_\mrm{ei},\\
\label{eq:two_fluid2}
\nabla \cdot \left(
\frac{5}{2}nu_\parallel T_\mrm{e} \mathbf{b}
- \kappa_{\parallel \mrm{e}} \nabla_{\parallel} T_\mrm{e}
- \kappa_{\perp \mrm{e}} \nabla_{\perp} T_\mrm{e}
\right)
&=-S_\mrm{ei}.
\end{align}
Here, $u_\parallel$ is the parallel flow velocity, $\mathbf{b}$ is the unit vector along the magnetic field, $T_\mrm{e}$ and $T_\mrm{i}$ are the electron and ion temperatures, and $S_\mrm{ei}$ is the electron--ion energy exchange term.
Note that equations~\ref{eq:two_fluid1} and \ref{eq:two_fluid2} neglect radiation and charge-exchange losses. This approximation is usually acceptable for attached plasmas in small and medium-sized devices, but it should be reconsidered for detached plasmas or larger devices, where SOL radiation becomes significant. In the SOL, the electron and ion temperatures are often comparable because of the high collisionality and the resulting thermal coupling \cite{brunner2013assessment}. We therefore assume $T_\mrm{i}=T_\mrm{e}=T$. Since the parallel ion heat conductivity is much smaller than the electron one, $\kappa_{\parallel \mrm{i}}\ll\kappa_{\parallel \mrm{e}}$ \cite{braginskii1965transport}, we set $\kappa_{\parallel \mrm{i}}\approx0$. Defining the total perpendicular conductivity as
$\kappa_{\perp}=\kappa_{\perp \mrm{i}}+\kappa_{\perp \mrm{e}}$,
the sum of equations \ref{eq:two_fluid1} and \ref{eq:two_fluid2} gives
\begin{align}
\label{eq:one_fluid}
\nabla \cdot \left(
5nu_\parallel T \mathbf{b}
- \kappa_{\parallel \mrm{e}} \nabla_{\parallel} T
- \kappa_{\perp} \nabla_{\perp} T
\right)
= 0 .
\end{align}

The relative importance of the terms in equation \ref{eq:one_fluid} depends on the SOL transport regime. In the sheath-limited regime, the plasma is nearly isothermal along the magnetic field lines, so that $\nabla_\parallel T\approx0$ \cite{stangeby2000plasma}. The parallel conductive term can then be neglected, and the parallel exhaust is mainly governed by convective transport to the sheath. On the other hand, in the conduction-limited regime, the temperature shows a maximum upstream and decreases towards the target. Parallel heat conduction then plays an important role, with $\kappa_{\parallel \mrm{e}}\propto T^{5/2}$ given by the Spitzer--Härm conductivity \cite{PhysRev.89.977}, and neither the parallel conduction nor the convective term can in general be neglected. TJ-II is known to be predominantly sheath-limited, as is typically the case for limiter SOLs in medium-sized devices \cite{stangeby2000plasma}. Therefore, we can adopt the approximation $\kappa_{\parallel e}\nabla_\parallel T\approx0$ in the present study.\\

\begin{figure*}
    \centering
    \includegraphics[width=1\linewidth]{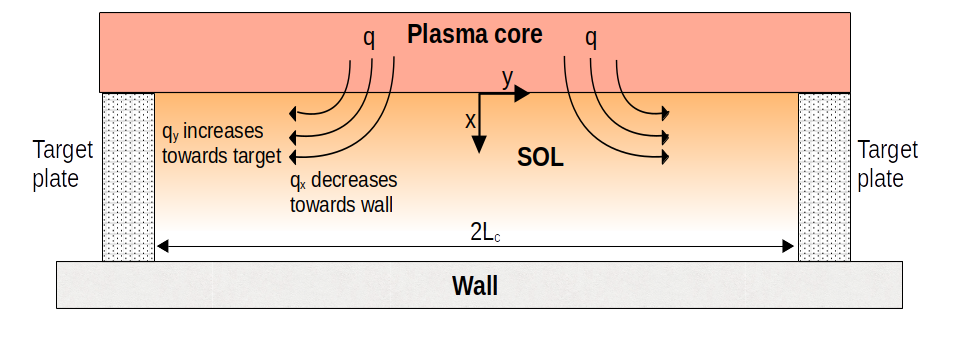}
    \caption{Schematic representation of the straightened SOL geometry used to derive the reduced transport model. The SOL is bounded by the two target plates separated by two times the connection length $L_\mrm{c}$. The coordinates $x$ and $y$ denote the perpendicular and parallel directions, respectively, with the origin of coordinates at the upstream position of the Last-Closed Flux Surface. Heat flux $\mathbf{q}$ enters the SOL through the $x$ direction, to be then redirected along the magnetic field lines and exhausted toward the targets.}
    \label{fig:SOL_scheme}
\end{figure*}

Strictly speaking, the divergence and gradient operators in equation~\ref{eq:one_fluid} should be evaluated in toroidal geometry. However, the SOL width is much smaller than the plasma minor radius, $\lambda_q\ll a$. Therefore, the SOL can be straightened out \cite[p.~20]{stangeby2000plasma}, so that the perpendicular and parallel directions define a 2D Cartesian basis, as illustrated in figure~\ref{fig:SOL_scheme}. Writing equation~\ref{eq:one_fluid} in this new basis for the sheath-limited case gives
\begin{align}
\label{eq:sheath_straight}
5T\frac{\partial\left( nu_y \right)}{\partial y}
- \frac{\partial}{\partial x}\left(\kappa_{\perp} \frac{\partial T}{\partial x}\right)
= 0,
\end{align}
where $x$ denotes now the perpendicular direction and $y$ the parallel direction.\\

The first term in equation~\ref{eq:sheath_straight} represents the divergence of the parallel convective heat flux. In a reduced 1D description along the perpendicular direction, this term can be interpreted as a local sink of thermal energy. Physically, the energy transported from the confined plasma into the SOL is progressively redirected along the magnetic field lines and exhausted at the target, so that the perpendicular heat flux decreases along the radial direction until it vanishes, as illustrated in figure~\ref{fig:SOL_scheme}.\\

To express this parallel loss in local form, we approximate the parallel particle flux term, $nu_y$, by using characteristic upstream and target quantities. At the upstream position, the net parallel particle flux is zero, since particles are lost symmetrically toward the two target ends. At the sheath entrance, the downstream density is approximately $n_\mrm{d}\approx n_\mrm{up}/2$, and the parallel velocity $u_y\approx c_\mrm{i}$ \cite{stangeby2000plasma}, where
\begin{equation}
c_\mrm{i}=\sqrt{\frac{T}{m_\mrm{i}}}
\end{equation}
is the ion sound speed, and $m_\mrm{i}$ is the ion mass. Assuming that the parallel particle flux varies linearly along $y$, its derivative can be estimated as
\begin{equation}
\label{eq:parallel_sink}
\frac{\partial\left( n u_y \right)}{\partial y}
\approx
\frac{n_\mrm{d}c_\mrm{i}-0}{L_\mrm{c}}
\approx
\frac{n_\mrm{up}c_\mrm{i}}{2L_\mrm{c}} ,
\end{equation}

where $L_\mrm{c}$ is the upstream-to-target connection length. Substituting equation \ref{eq:parallel_sink} into equation \ref{eq:sheath_straight} gives the reduced 1D perpendicular transport equation
\begin{equation}
\label{eq:1D_eq}
\frac{\partial}{\partial x}
\left(
\kappa_{\perp} \frac{\partial T}{\partial x}
\right)
=
\frac{5n_\mrm{up}T^{3/2}}{2\sqrt{m_\mrm{i}}L_\mrm{c}} ,
\end{equation}

that we take to model the SOL region of seath-limited devices. It should be noted that equation 7 is a strongly simplified description of the heat transport in the SOL, possibly relevant for the conditions found in small devices such as TJ-II. However, the method presented in this work can be adapted to more sophisticated descriptions.\\

It is also necessary to specify the boundary condition at the interface between the confined and unconfined plasma, namely the Last-Closed Flux Surface (LCFS), located at $x=0$ in the present coordinate system. At this boundary, the transported power $P_\mrm{tr}$ leaving the confined plasma is assumed to enter the SOL uniformly over the total LCFS area, $S_\mrm{LCFS}$. This gives the following relation between the perpendicular heat flux, the perpendicular conductivity, the temperature gradient, and the transported power:
\begin{equation}
\label{eq:LCFS_heat_flux}
q_x^\mrm{LCFS}
=
-\left.
\kappa_\perp \frac{\partial T}{\partial x}
\right|_{x_\mrm{LCFS}}
=
\frac{P_\mrm{tr}}{S_\mrm{LCFS}} .
\end{equation}

\subsubsection{Existing \texorpdfstring{$\kappa_\perp$}{kappa-perp} scalings}
\label{subsubsec:kperp_scalings}

Here, we summarize the main scaling laws commonly used to describe perpendicular transport coefficients in the confined region of fusion plasmas. Although these scalings were not originally derived for SOL transport, they provide useful reference trends for interpreting the results obtained in section~\ref{sec:exp}, allowing us to assess whether the inferred $\kappa_\perp(n,T)$ shows trends consistent with established plasma transport regimes.

\begin{itemize}
  \item Classical and neoclassical transport: $\kappa_\perp\propto \frac{n^2}{B^2\sqrt{T}}$ \cite[p.~217]{braginskii1965transport}\cite{helander2012classical}. Both Braginskii (classical) and  Pfirsch–Schlüter (neoclassical) coefficients show the same dependence on $T$, $n$ and $B$, although the Pfirsch--Schlüter coefficient includes the enhancement of perpendicular transport caused by toroidal geometry. Both descriptions are derived for collisional, fully ionized plasmas.
  \item Bohm transport: $\kappa_\perp\propto \frac{nT}{B}$ \cite{bohm1949characteristics}. Bohm transport was first introduced empirically by D. Bohm in 1949. It is usually interpreted as an anomalous transport scaling in which the step size does not depend on the gyroradius $\rho_L$ but on the thermal velocity of particles.
  \item Gyro-Bohm transport: $\kappa_\perp\propto \frac{nT^{3/2}}{B^2}$ \cite{manfredi1997gyro}. In contrast to Bohm transport, Gyro-Bohm transport assumes that turbulent perpendicular transport is controlled by $\rho_L$. It can be obtained by multiplying the Bohm diffusivity by $\rho_L/a$, where $a$ is the plasma minor radius.
\end{itemize}

The ISS04 energy confinement time scaling is compatible with gyro-Bohm diffusivity \cite{yamada2005characterization}, and therefore core perpendicular transport in stellarators is often considered to follow gyro-Bohm-like behaviour. However, it is not clear that this behaviour can be extrapolated to the SOL, where open field lines, lower temperature and higher collisionality can substantially modify the transport regime.

\subsection{PINN inverse problem}
\label{subsec:inverse_pinns}

The inverse PINN formulation provides a way to infer unknown parameters in a differential equation by combining sparse measurements with the physical constraints imposed by the governing equation. In contrast to a forward problem, where the coefficients of the equation are known and the solution is sought, the inverse problem aims to recover one or more unknown coefficients from partial observations of the solution. This section introduces the formulation progressively. First, the case of an unknown constant coefficient is considered, establishing the basic structure of the inverse PINN loss function. The formulation is then extended to the case in which the unknown coefficient is not a scalar, but a function of the solution variable itself.

\subsubsection{Constant $\kappa$}

Consider a differential equation of the form
\begin{equation}
    \mathscr{N}\left[U(x);\;\kappa\right]=0,
    \qquad x\in \Omega \subset \mathbb{R},
\end{equation}
where $\mathscr{N}$ is a differential operator, $U$ is the dependent variable, $x$ is the independent variable, and $\kappa$ is an unknown constant parameter. In the inverse PINN formulation, the solution $U$ is approximated by a neural network $\hat{U}(x;\Theta_U)$, where $\Theta_U$ denotes the network weights. The unknown parameter is represented by an additional trainable scalar, $\hat{\kappa}$. The physics residual is then defined as
\begin{equation}
\label{eq:residual_pinn_constant}
    f(x;\Theta_U,\hat{\kappa})
    :=
    \mathscr{N}\left[\hat{U}(x;\Theta_U);\;\hat{\kappa}\right].
\end{equation}
This residual defines the physics-informed part of the neural-network model \cite{raissi2019physics}. The network weights $\Theta_U$ and the parameter $\hat{\kappa}$ are learned simultaneously by minimizing a loss function that combines the residual of the differential equation with the mismatch between the neural network solution and the available measurements:
\begin{equation}
\label{eq:loss_pinn_constant}
    L =
    \frac{1}{N_f}\sum_{i=1}^{N_f}
    \left|f(x_f^i;\Theta_U,\hat{\kappa})\right|^2
    +
    \frac{1}{N_U}\sum_{i=1}^{N_U}
    \left|U^i-\hat{U}(x_U^i;\Theta_U)\right|^2 .
\end{equation}
Here, $\{x_U^i,U^i\}_{i=1}^{N_U}$ denotes the labelled dataset, while $\{x_f^i\}_{i=1}^{N_f}$ denotes the collocation points at which the residual of the differential equation is evaluated. The number of available measurements, $N_U$, is determined by the experiment and is often scarce. In contrast, the number of collocation points, $N_f$, can be sampled freely during training, subject to computational cost, and is therefore treated as a hyperparameter. Both the measurement points $\{x_U^i\}$ and the collocation points $\{x_f^i\}$ should cover the domain $\Omega$ sufficiently well to constrain the solution.\\

\subsubsection{Functional $\kappa(U)$}\label{subsubsec:functional_kappa}

The inverse formulation can be extended to the case in which the unknown coefficient is not a constant, but a function of the solution variable, $\kappa=\kappa(U)$. In this case, $\kappa(U)$ is represented by a second neural network, $\hat{\kappa}(U;\Theta_\kappa)$. The residual becomes
\begin{equation}
\label{eq:residual_pinn_functional}
    f(x;\Theta_U,\Theta_\kappa)
    :=
    \mathscr{N}
    \left[
    \hat{U}(x;\Theta_U);
    \;\hat{\kappa}\left(\hat{U}(x;\Theta_U);\Theta_\kappa\right)
    \right] .
\end{equation}
Note that, during training, the input to the $\kappa$ network is the solution predicted by $\hat{U}$. The weights $\Theta_U$ and $\Theta_\kappa$ are learned simultaneously, so that both the reconstructed solution and the inferred coefficient are consistent with the measurements and with the governing equation.\\

Allowing $\kappa$ to be an arbitrary function introduces an additional degree of freedom into the inverse problem. Consequently, the available measurements of $U$ and the differential equation may not be sufficient to identify a unique $\kappa(U)$. To avoid this degeneracy, additional information on $\kappa$ can be imposed, for example its value at a boundary or at selected reference points. The loss function can then be written as
\begin{align}
\label{eq:loss_pinn_functional}
    L =&
    \frac{1}{N_f}\sum_{i=1}^{N_f}
    \left|f(x_f^i;\Theta_U,\Theta_\kappa)\right|^2 \nonumber\\ &
    +\frac{1}{N_U}\sum_{i=1}^{N_U}
    \left|U^i-\hat{U}(x_U^i;\Theta_U)\right|^2
    \nonumber\\ 
    &+\frac{1}{N_\kappa}\sum_{i=1}^{N_\kappa}
    \left|\kappa^i-\hat{\kappa}(U_\kappa^i;\Theta_\kappa)\right|^2 .
\end{align}
The last term enforces the available constraints on the unknown coefficient, where $\{U_\kappa^i,\kappa^i\}_{i=1}^{N_\kappa}$ are reference values used to remove the non-uniqueness of the inverse problem.\\

\section{PINN model}
\label{sec:PINN_model}
The objective of the PINN model is to infer the perpendicular heat conductivity function, $\kappa_\perp(n,T)$, while simultaneously reconstructing the temperature and density profiles from sparse measurements. The inverse PINN model used in this work was implemented in Python using NeuroDiffEq \cite{chen2020neurodiffeq,liu2025recent}, an open-source library for Physics-Informed Neural Network applications built on top of PyTorch \cite{paszke2019pytorch}. This framework provides the automatic differentiation tools required to evaluate the derivatives appearing in the transport equation residual, together with the neural network training infrastructure used to optimize the model parameters.\\

This section first describes the architecture of the PINN model. The optimization procedure is then presented, including the update of the neural network parameters and the adaptive loss weights used to balance the different contributions to the total loss.

\subsection{Proposed PINN architecture}
\label{subsec:pinn_architecture}

\begin{figure*}
    \centering
    \includegraphics[width=1.05\linewidth]{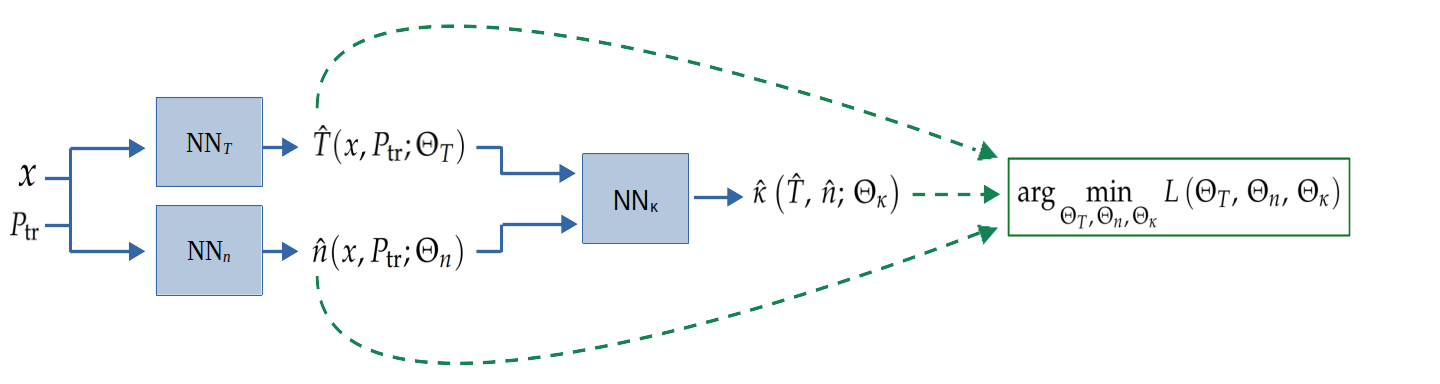}
    \caption{Schematic representation of the inverse PINN architecture used to infer $\kappa_\perp(n,T)$. The temperature and density profiles are computed by independent ResNet neural networks from the perpendicular coordinate $x$ and the transported power $P_\mrm{tr}$, while the inferred profiles are used as inputs to a third ResNet network representing the perpendicular heat conductivity. The three networks are trained simultaneously by minimizing the data losses and the residual of the 1D transport equation.}
    \label{fig:scheme_PINN}
\end{figure*}

The physical constraint embedded in the loss function is the reduced 1D transport equation derived in equation~\ref{eq:1D_eq}. In the following, the measurements are assumed to be taken sufficiently close to the upstream region, so that the subscript in $n_\mrm{up}$ is dropped for simplicity. We rewrite equation~\ref{eq:1D_eq} to show explicitly the relevant dependencies:
\begin{equation}
\label{eq:1D_eq_pinn}
\frac{\partial}{\partial x}
\left[
\kappa_{\perp}(n,T) \frac{\partial T(x)}{\partial x}
\right]
-
\frac{5\,n(x)[T(x)]^{3/2}}{2\sqrt{m_\mrm{i}}L_\mrm{c}}
=0 .
\end{equation}
That is, $T=T(x)$, $n=n(x)$, $\kappa_\perp=\kappa_\perp(n,T)$.\\

Following the functional formulation introduced in section~\ref{subsubsec:functional_kappa}, these three quantities are approximated by neural networks. A first possibility would be to represent the plasma profiles only as functions of $x$: $\hat{T}(x)$ and $\hat{n}(x)$. However, this would restrict the training to a single plasma scenario, corresponding to a given transported power $P_\mrm{tr}$. Since the SOL temperature profile depends on the power crossing the LCFS, as expressed through the boundary condition in equation~\ref{eq:LCFS_heat_flux}, the transported power $P_\mrm{tr}$ is introduced as an additional input, as shown in figure~\ref{fig:scheme_PINN}. This follows the bundle method idea first described in Flamant \textit{et al.}~\cite{flamant2020solving}. Hence, the temperature and density networks are expressed as
\begin{equation}
    \hat{T}
    =
    \hat{T}(x,P_\mrm{tr};\Theta_T),
    \qquad
    \hat{n}
    =
    \hat{n}(x,P_\mrm{tr};\Theta_n),
\end{equation}
where $\Theta_T$ and $\Theta_n$ are the corresponding trainable parameters. Both networks are built using ResNet blocks \cite{he2016deep} to improve gradient propagation and facilitate the training of deeper architectures. In this formulation, $P_\mrm{tr}$ labels different plasma profiles obtained under different heating conditions. This allows a single PINN model to be trained simultaneously on several discharges, provided that the magnetic configuration is kept fixed.\\

The predicted profiles are then passed to a third ResNet neural network, which represents the perpendicular conductivity as a function of the temperature and density:
\begin{equation}
    \hat{\kappa}_\perp
    =
    \hat{\kappa}_\perp
    \left(
    \hat{T},\hat{n};\Theta_\kappa
    \right),
\end{equation}
where $\Theta_\kappa$ denotes the trainable parameters of the conductivity network. The three networks are trained simultaneously, so that the reconstructed profiles fit the labelled measurements while the inferred conductivity makes them consistent with the transport equation.\\

The physics residual is obtained by substituting the neural network approximations into equation~\ref{eq:1D_eq_pinn}:
\begin{align}
\label{eq:pinn_residual_sol}
&f(x,P_\mrm{tr};\Theta_T,\Theta_n,\Theta_\kappa)
=\nonumber\\
&\frac{\partial}{\partial x}
\left[
\hat{\kappa}_{\perp}
\left(
\hat{T},\hat{n};\Theta_\kappa
\right)
\frac{\partial \hat{T}(x,P_\mrm{tr};\Theta_T)}{\partial x}
\right]
-\nonumber\\
&\frac{5\,\hat{n}(x,P_\mrm{tr};\Theta_n)[\hat{T}(x,P_\mrm{tr};\Theta_T)]^{3/2}}
{2\sqrt{m_\mrm{i}}L_\mrm{c}} .
\end{align}

The total loss function is defined as the weighted sum of four partial losses:
\begin{equation}
\label{eq:total_loss_pinn}
    L
    =
    \lambda_f L_f
    +
    \lambda_T L_T
    +
    \lambda_n L_n
    +
    \lambda_\kappa L_\kappa ,
\end{equation}
where $L_f$ is the physics loss, $L_T$ and $L_n$ are the data losses for temperature and density, and $L_\kappa$ imposes the boundary condition on the conductivity. These terms have different physical units and numerical scales, so their relative contribution to the total loss may differ substantially. To prevent any single term from dominating the optimization, the coefficients $\lambda_f$, $\lambda_T$, $\lambda_n$, and $\lambda_\kappa$ are introduced as partial loss weights. Their purpose is to balance the contribution of the four partial losses throughout training. Their update rule together with the training procedure will be described in section~\ref{subsec:Network_optimization}.\\

Each loss term is computed as a mean-squared error:
\begin{align}
\label{eq:loss_terms_pinn}
    L_f
    &=
    \frac{1}{N_f}\sum_{i=1}^{N_f}
    \left|
    f(x_f^i,P_{\mrm{tr},f}^i;\Theta_T,\Theta_n,\Theta_\kappa)
    \right|^2 ,
    \\
    L_T
    &=
    \frac{1}{N_D}\sum_{i=1}^{N_D}
    \left|
    T^i
    -
    \hat{T}(x_D^i,P_{\mrm{tr},D}^i;\Theta_T)
    \right|^2 ,
    \\
    L_n
    &=
    \frac{1}{N_D}\sum_{i=1}^{N_D}
    \left|
    n^i
    -
    \hat{n}(x_D^i,P_{\mrm{tr},D}^i;\Theta_n)
    \right|^2 ,
    \\
    L_\kappa
    &=
    \frac{1}{N_\kappa}\sum_{i=1}^{N_\kappa}
    \left|
    \kappa_\perp^i
    -
    \hat{\kappa}_\perp(T_\kappa^i,n_\kappa^i;\Theta_\kappa)
    \right|^2 .
\end{align}
The set $\{x_f^i,P_{\mrm{tr},f}^i\}_{i=1}^{N_f}$ denotes the collocation points at which the residual of the transport equation is evaluated. The set $\{x_D^i,P_{\mrm{tr},D}^i,T^i,n^i\}_{i=1}^{N_D}$ denotes the labelled dataset of temperature and density measurements at different radial positions and power conditions. Finally, $\{T_\kappa^i,n_\kappa^i,\kappa_\perp^i\}_{i=1}^{N_\kappa}$ denotes the reference values used to constrain the conductivity. Such values are imposed at the LCFS $(x=0)$, where equation~\ref{eq:LCFS_heat_flux} relates the perpendicular heat flux and therefore $\kappa_\perp$ to the transported power.

\subsection{Network optimization}
\label{subsec:Network_optimization}

Once the loss function has been defined, the trainable parameters of the three neural networks,
$\Theta_T$, $\Theta_n$, and $\Theta_\kappa$, are optimized simultaneously. The optimization problem is written as
\begin{equation}
\label{eq:optimization_problem}
\Theta_T^\ast,\Theta_n^\ast,\Theta_\kappa^\ast
=
\argmin_{\Theta_T,\Theta_n,\Theta_\kappa}
L(\Theta_T,\Theta_n,\Theta_\kappa),
\end{equation}
where $L$ is the total loss defined in equation~\ref{eq:total_loss_pinn}, and
$\Theta_T^\ast$, $\Theta_n^\ast$, and $\Theta_\kappa^\ast$ denote the optimized network parameters after training. The minimization is performed using the Adam optimizer \cite{kingma2014adam}, a stochastic gradient-based method that adaptively updates each parameter using estimates of the first and second moments of the gradients. The gradients are computed by backpropagation using automatic differentiation, which provides both the derivatives required to evaluate the differential equation residual and the gradients to update the neural network parameters.\\

The weights assigned to the partial losses play an important role in the optimization. They serve two purposes: first, to bring the different partial losses to comparable numerical scales at the beginning of training; and second, to dynamically balance their contribution during training according to the magnitude of their backpropagated gradients. The loss weights are defined as
\begin{align}
    \lambda_f &= \alpha_f,\\
    \lambda_j &= \alpha_j\gamma_j,
\end{align}
where $j\in\{T,n,\kappa\}$. The coefficients $\alpha_j$ provide the initial normalization of the partial losses. They are computed from the losses evaluated at epoch zero as
\begin{align}
\label{eq:lambda_fix}
    \alpha_f &= \frac{1}{L_f^{(0)}},\\
    \alpha_j &= \frac{1}{L_j^{(0)}},
\end{align}
where the superscript $(0)$ denotes evaluation before training. This normalization initializes the contribution of each partial loss to be all equally 1.\\

The coefficients $\gamma_j$ are then updated dynamically during training every $l_\gamma$ epochs, following the balancing strategy proposed by Deguchi \textit{et al}. \cite{deguchi2023dynamic}. At each update index $k$, an instantaneous correction factor is computed as
\begin{equation}
\label{eq:gamma_hat}
    \hat{\gamma}_j^{(k)}
    =
    \frac{
    \left\lVert\nabla_\Theta \left(\alpha_fL_f^{(k)}\right)\right\rVert_2
    }{
    \left\lVert\nabla_\Theta\left(\alpha_jL_j^{(k)}\right)\right\rVert_2
    },
\end{equation}
where $\nabla_\Theta$ denotes the gradient with respect to all the trainable network parameters. The adaptive coefficient is then updated using an exponential moving average:
\begin{equation}
\label{eq:exp_average}
    \gamma_j^{(k)}
    =
    \beta\gamma_j^{(k-1)}
    +
    \left(1-\beta\right)\hat{\gamma}_j^{(k)} ,
\end{equation}
where $\beta\in[0,1)$ controls the smoothing of the update, and $\gamma_j^{(0)}$ is initialized to 1.
In this formulation, the physics loss is taken as the reference term and therefore no additional dynamic coefficient $\gamma_f$ is introduced for it. The adaptive factors $\gamma_j$ rescale the data and conductivity losses so that their backpropagated gradient norms remain comparable to that of the physics residual. The exponential averaging in equation~\ref{eq:exp_average} reduces fluctuations in the weights and prevents noisy updates from destabilizing the training.

\section{Validation with synthetic data}
\label{sec:synthetic_validation}

\begin{figure*}
    \centering
    \includegraphics[width=1\linewidth]{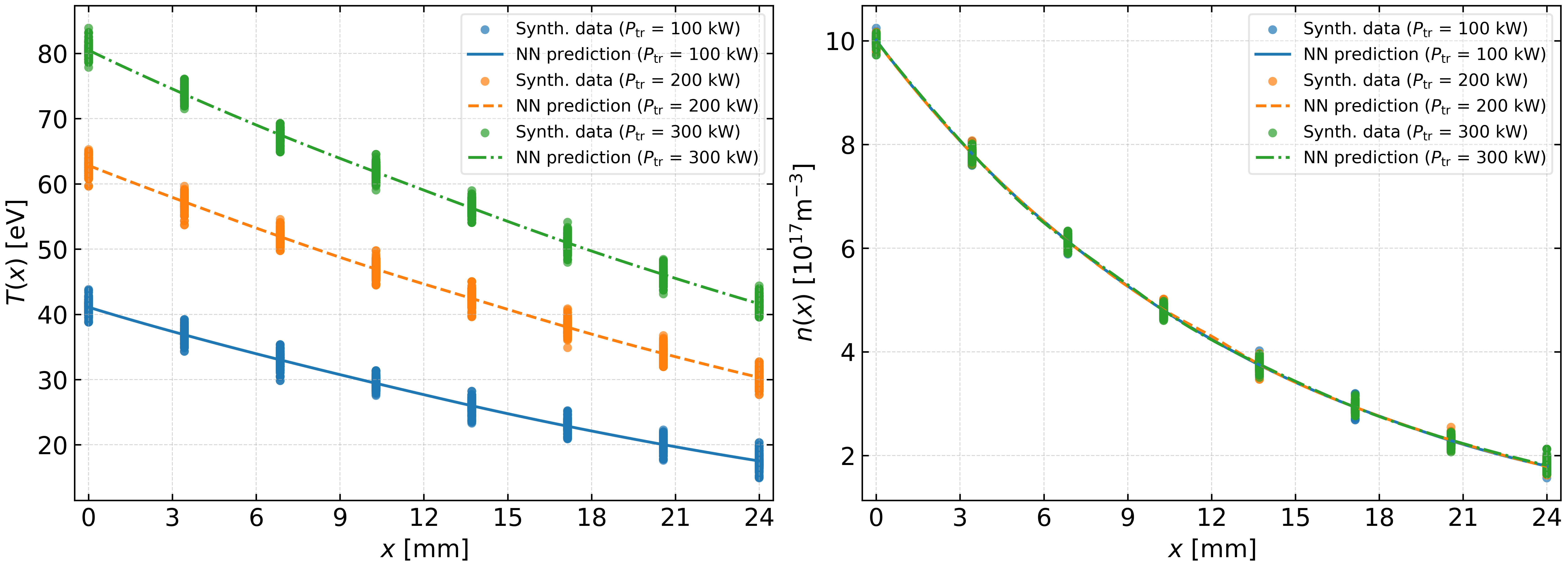}
    \caption{Comparison between synthetic data and neural network reconstructions for a representative use case of the model. Left: temperature profiles $T(x)$ for three transported power values. Right: density profiles $n(x)$ for the same cases. Markers correspond to noisy synthetic measurements sampled at discrete radial positions, while lines represent the smooth profiles reconstructed by the neural networks.}
    \label{fig:T_n_synth}
\end{figure*}

The PINN model is first validated using synthetic data. This step provides a controlled test case in which the density data and the functional form of the perpendicular conductivity are prescribed a priori, while the temperature data are obtained by solving the transport equation numerically, as described in section \ref{subsec:synthetic_data_generation}. The inferred conductivity, $\hat{\kappa}_\perp(n,T)$, can therefore be compared directly with the ground-truth function $\kappa_\perp(n,T)$ used to generate the data. In section \ref{subsec:numerical_results}, we first present a representative application of the model. We then perform a two dimensional scan as a function of the number of profiles processed simultaneously and the number of radial positions sampled per profile. This scan allows us to assess how the PINN performance depends on the amount of available data, and whether the reconstruction accuracy saturates beyond a certain amount of data.

\subsection{Synthetic data generation}
\label{subsec:synthetic_data_generation}

The density profile is prescribed as an exponentially decaying function,
\begin{equation}
    n(x)=n_0\exp\left(-\frac{x}{W}\right),
\end{equation}
where $n_0=10^{18}\:\mrm{m^{-3}}$ is a typical density value at the LCFS, and $W=14\:\mrm{mm}$ characterizes the radial decay length of the density profile. The perpendicular conductivity is chosen as
\begin{equation}\label{eq:ground_kappa}
    \kappa_\perp(n,T)=k_0 n T,
\end{equation}
where $k_0=1\:\mrm{m^2s^{-1}eV^{-1}}$. This imposed conductivity is not intended to represent a definitive SOL transport scaling. Instead, it provides a known dependence on both density and temperature that remains physically reasonable in light of current understanding of SOL transport, allowing the inverse method to be quantitatively assessed.\\

To generate the synthetic temperature profiles, we introduce the transient relaxation equation associated with the steady transport balance in equation~\ref{eq:1D_eq_pinn}:
\begin{equation}
\label{eq:time_dependent_heat_transport}
n\frac{\partial T}{\partial t}
=
\frac{\partial}{\partial x}
\left[
\kappa_\perp
\frac{\partial T}{\partial x}
\right]
-
\frac{5nT^{3/2}}{2\sqrt{m_\mrm{i}}L_\mrm{c}} ,
\end{equation}
For each transported power value, equation~\ref{eq:time_dependent_heat_transport} is solved subject to the LCFS heat-flux boundary condition in equation~\ref{eq:LCFS_heat_flux} for $x_\mrm{init}=0$. The outer boundary of the computational domain is placed sufficiently far from the LCFS, at $x_\mrm{end}=140~\mrm{mm}$, such that the temperature decays to approximately zero before reaching it. Consequently, the solution in the region of interest is insensitive to the specific boundary condition imposed at $x=x_\mrm{end}$. Non-negative temperatures, $T(x,t)\geq0$, are enforced throughout the simulation. The LCFS area entering equation~\ref{eq:LCFS_heat_flux} is approximated as
\begin{equation}
    S_\mrm{LCFS}\approx4\pi^2aR,
\end{equation}
where $a$ and $R$ are the minor and major radii of the device, respectively. For TJ-II, this gives $S_\mrm{LCFS}\approx11.8\:\mrm{m^2}$.\\

Equation~\ref{eq:time_dependent_heat_transport} is solved in Python using the method of lines. Spatial derivatives are approximated by finite differences on a uniform grid spanning $x\in[0,140]~\mrm{mm}$ with spacing $\Delta x=0.14~\mrm{mm}$, yielding a system of ordinary differential equations that is advanced with the implicit Radau solver implemented in the \texttt{solve\_ivp} routine from SciPy \cite{virtanen2020scipy}. The Radau solver controls the local error $\epsilon$ according to the criterion $\epsilon <a_\mrm{tol}+r_\mrm{tol}|T|$, with relative and absolute tolerances set to $r_\mrm{tol}=10^{-3}$ and $a_\mrm{tol}=10^{-6}$, respectively. The system is integrated up to $t_\mrm{end}=0.05~\mrm{s}$, at which point a stationary state is reached. Specifically, the time derivative term averaged over the spatial domain is ten orders of magnitude smaller than that of each of the two terms on the right-hand side of equation~\ref{eq:time_dependent_heat_transport}. In this limit, equation~\ref{eq:time_dependent_heat_transport} reduces to equation~\ref{eq:1D_eq_pinn}, and the resulting stationary temperature profile is used as synthetic data.\\

The training dataset, $\{x_D^i,P_{\mrm{tr},D}^i,T^i,n^i\}_{i=1}^{N_D}$, is then obtained by sampling the stationary profiles at equally-spaced discrete values of $x_D^i$ and $P_{\mrm{tr},D}^i$. The selected ranges, $x_D^i\in [0,24]\:\mrm{mm}$ and $P_{\mrm{tr},D}^i\in [100,300]\:\mrm{kW}$, are consistent with values typically observed in TJ-II \cite{ivanova2024characterization}. Figure~\ref{fig:T_n_synth} shows an example synthetic dataset with $N_\mrm{P}=3$ distinct values of $P_{\mrm{tr},D}$ and $N_\mrm{x}=8$ distinct radial positions. Markers indicate the sampled synthetic data. For each unique radial position, $N_\mrm{R}$ repeated measurements of $T$ and $n$ are generated by adding independent Gaussian noise to the original synthetic data. This procedure mimics the experimental acquisition process, in which temperature and density are measured at a given time resolution and are affected by plasma fluctuations. Assuming a sampling frequency of $2~\mrm{kHz}$, consistent with Langmuir probe diagnostics, and a discharge duration of $100~\mrm{ms}$, we set $N_\mrm{R}=200$. The standard deviations of the Gaussian noise are chosen as $\sigma_\mrm{T}=1~\mrm{eV}$ for the temperature and $\sigma_\mrm{n}=10^{17}~\mrm{m^{-3}}$ for the density, simulating typically observed measurement errors and plasma fluctuations \cite{ivanova2024characterization}.\\

The conductivity data at the LCFS, $\{\kappa_\perp^i\}_{i=1}^{N\kappa}$, are computed from equation~\ref{eq:ground_kappa} using the noiseless synthetic temperature and density values at the LCFS. One conductivity value at the LCFS is assigned to each distinct transported power case; therefore, $N_\kappa=N_\mrm{P}$. For the example shown in figure~\ref{fig:T_n_synth}, this yields three LCFS conductivity values, corresponding to the three transported power profiles.

\subsection{Numerical results}\label{subsec:numerical_results}
This section first presents a representative training run to illustrate the general behaviour of the PINN inverse problem solver and the reconstruction of $\kappa_\perp(n,T)$. A bootstrap resampling procedure is then introduced to estimate the uncertainty and reproducibility of the inferred conductivity. Finally, a systematic scan is performed over the number of input profiles, $N_\mrm{P}$, and the number of radial measurement positions per profile, $N_\mrm{x}$. The hyperparameters listed in table~\ref{tab:hyperparameters} are kept fixed throughout all numerical experiments. The collocation set is constructed as the Cartesian product of $256$ $x_f$ values and $64$ $P_{\mrm{tr},f}$ values, resulting in $N_f=256\times64=16384$ collocation points.\\

\begin{table*}
    \centering
    \caption{Training hyperparameters used in the PINN model.}
    \label{tab:hyperparameters}
    \begin{tabular}{p{0.70\textwidth} p{0.20\textwidth}}
        \toprule
        \textbf{Hyperparameter} & \textbf{Value} \\
        \midrule
        Total number of training epochs. & $5\cdot10^{5}$ \\
        Learning rate used by the optimizer. & $10^{-4}$ \\
        Number of neurons in each hidden layer of the neural networks. & $128$ \\
        Number of ResNet blocks in the neural networks. & $5$ \\
        Activation function used in the hidden layers. & \texttt{ELU} \\
        Activation function used in output layer. & \texttt{Identity} \\
        Exponential smoothing factor used in the dynamic loss weight update $(\beta)$. & $0.999$ \\
        Update interval used in the dynamic loss weight update $(l_\gamma)$. & $10$ \\
        Number of generated collocation points in the spatial coordinate. & $256$ \\
        Number of generated collocation points in the power coordinate. & $64$ \\
        Number of training batches evaluated per epoch. & $1$ \\
        Random seed used for stochastic initialization and sampling. & Not fixed\\
        \bottomrule
    \end{tabular}
\end{table*}

\subsubsection{General example}
\begin{figure}[t]
    \centering
    \includegraphics[width=1\linewidth]{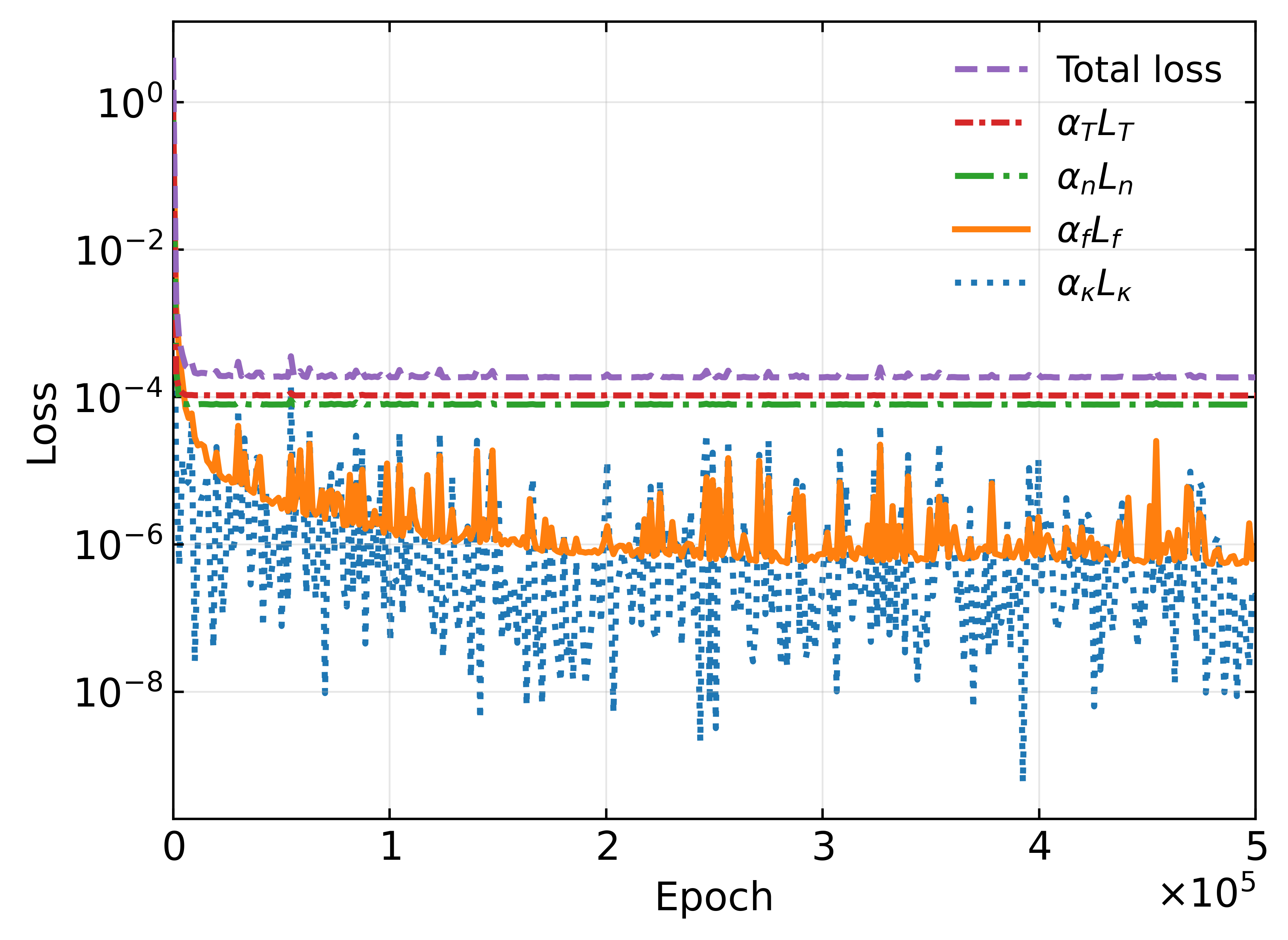}
    \caption{Evolution of the individual partial losses normalized by the static coefficients $\alpha_j$, and the resulting total loss, during the representative synthetic data run. The losses are shown in logarithmic scale as functions of the training epoch.}
    \label{fig:train_loss}
\end{figure}
We first apply the model to the synthetic data from figure~\ref{fig:T_n_synth}. Figure~\ref{fig:train_loss} shows the training loss evolution of all loss terms. The temperature and density data losses reach a higher plateau earlier than the $\kappa$ and $f$ losses. This plateau represents an irreducible error, or noise floor, caused by the added measurement noise: as shown in figure~\ref{fig:T_n_synth}, the predicted profiles $\hat{T}$ and $\hat{n}$ follow the mean trend of each data cluster but cannot exactly interpolate every noisy data point, so further fitting or training cannot eliminate this residual. Despite this limitation, the optimization converges to a stable loss plateau. The model requires approximately 3 hours to converge when trained on the Turgalium HPC cluster using an NVIDIA H100 GPU.
 \\

We define the normalized physics-consistency score $\mathcal{S}_f$ as
\begin{equation}
\label{eq:S_DE}
    \mathcal{S}_f
    =
    1-
    \frac{
    \sum_{i=1}^{N_\mrm{D}} |f^i|
    }{
    \sum_{i=1}^{N_\mrm{D}}
    \left(
    |\mathcal{L}^i|+|\mathcal{R}^i|
    \right)
    },
\end{equation}
Here, $\mathcal{L}^i$ and $\mathcal{R}^i$ denote the first and second physical terms of equation~\ref{eq:pinn_residual_sol}, respectively, evaluated at the $i$-th data point, and the local residual is $f^i=\mathcal{L}^i-\mathcal{R}^i$. The normalized physics-consistency score $\mathcal{S}_f$, which ranges from 0 to 1, measures how well the differential transport equation is balanced by comparing the absolute local residual, $|f^i|$, with the combined magnitudes of its two physical terms, $|\mathcal{L}^i|+|\mathcal{R}^i|$. A value of $\mathcal{S}_f=1$ corresponds to exact balance, while lower values indicate greater normalized imbalance. This score is used only after training as a diagnostic and interpretive measure of physical consistency; it is not a training loss, stopping criterion, or model-selection quantity. For the present case, $\mathcal{S}_\mrm{f}=0.998$, indicating that the differential equation is nearly perfectly satisfied.\\

\begin{figure*}
    \centering
    \includegraphics[width=1\linewidth]{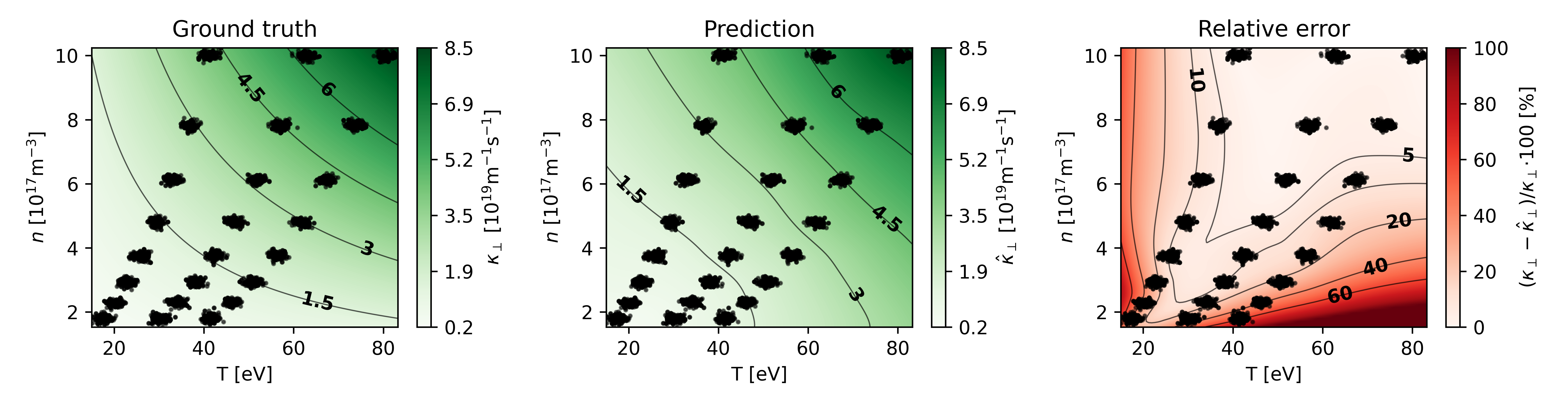}
    \caption{Comparison between the ground truth $\kappa_\perp$ and PINN-reconstructed $\hat{\kappa}_\perp$ perpendicular thermal conductivity as a function of temperature and density. The left panel shows the imposed ground truth conductivity (equation \ref{eq:ground_kappa}). The central panel shows the PINN-reconstructed conductivity. The right panel shows the relative error between $\kappa_\perp$ and $\hat{\kappa}_\perp$.  Black markers indicate the synthetic training dataset $\{T^i,\,n^i\}$.}
    \label{fig:kappa_dif}
\end{figure*}

The main quantity of interest is the inferred conductivity $\hat{\kappa}_\perp(T,n)$. Figure~\ref{fig:kappa_dif} compares $\hat{\kappa}_\perp(T,n)$ pointwise with the prescribed ground-truth conductivity $\kappa_\perp(T,n)$ defined in equation~\ref{eq:ground_kappa} over the $(T,n)$ domain. The right panel shows the percentage difference normalized by the ground-truth, in absolute values:
\begin{equation}
    \delta_\kappa(T,n)=\left|\frac{\kappa_\perp(T,n)-\hat{\kappa}_\perp(T,n)}{\kappa_\perp(T,n)}\right|\cdot 100.
\end{equation}
In the region constrained by the training dataset, the magnitude $\delta_\kappa$ is generally below $10\%$.\\

In the synthetic data case, the reconstruction error can be evaluated directly because the ground truth conductivity is known. However, this comparison is not possible when the framework is applied to experimental data. An alternative measure is therefore needed to evaluate the model consistency in predicting $\hat{\kappa}_\perp\left(T,n\right)$ without access to its true value. For this purpose, we use bootstrap resampling \cite{efron1992bootstrap}. The basic resampling unit is a single aligned observation, defined by the tuple $(T,n,x,P_\mrm{tr})$. All observations are pooled across transported power values, and bootstrap samples are generated by drawing these tuples with replacement from the complete dataset. Each bootstrap sample contains the same number of observations as the original dataset, preserving the underlying distribution while differing in the specific observations selected. This procedure generates alternative realizations of the empirical dataset that can be used to quantify the sensitivity of the inferred $\hat{\kappa}\perp(T,n)$ to variations in the sampled training data.

\begin{figure}
    \centering
    \includegraphics[width=1\linewidth]{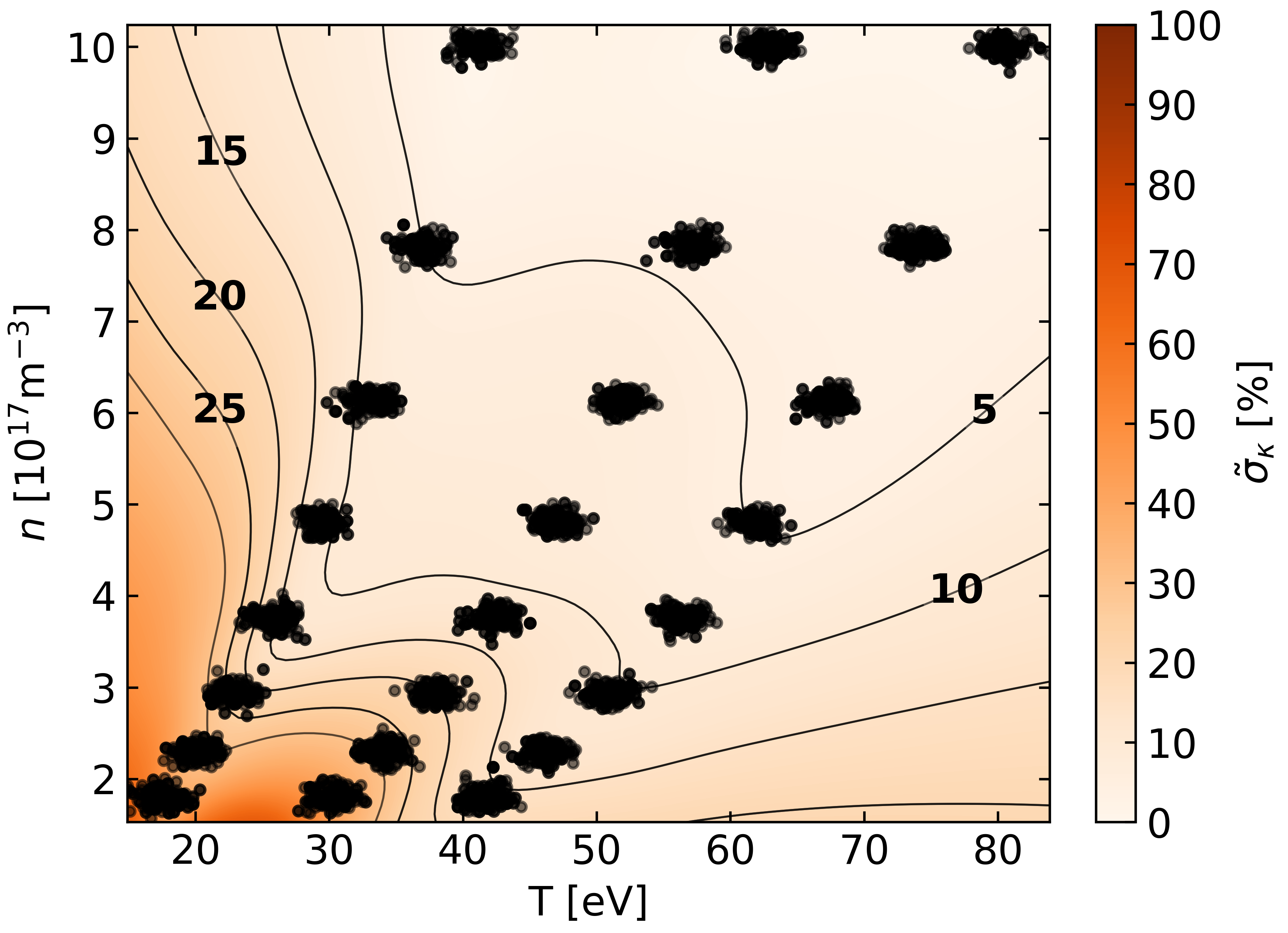}
    \caption{Pointwise standard deviation of the inferred perpendicular conductivity $\hat{\kappa}_\perp(T,n)$ obtained from 10 bootstrap realizations. $\tilde{\sigma}_\kappa$ is normalized by the mean $\overline{\kappa}_\perp(T,n)$, and given as a percentage. Black markers indicate the synthetic training dataset $\{T^i,\,n^i\}$.}
    \label{fig:kappa_bootstrap_std}
\end{figure}

The PINN is trained independently on each bootstrap sample, producing an ensemble of conductivity estimates, $\{\hat{\kappa}_\perp^{(b)}(T,n)\}_{b=1}^{N_\mrm{B}}$, where $N_\mrm{B}$ is the number of bootstrap realizations. The pointwise standard deviation of this ensemble can then be used to quantify the sensitivity to variations in the sampled measurements. Namely, figure~\ref{fig:kappa_bootstrap_std} shows $\tilde{\sigma}_\kappa$, defined as

\begin{equation}
\label{eq:kappa_bootstrap_std_norm}
\begin{aligned}
\tilde{\sigma}_\kappa(T,n)
&=
100\,
\frac{1}{\overline{\kappa}_\perp(T,n)}
\\
&\quad\times
\Biggl[
\frac{1}{N_\mrm{B}-1}
\sum_{b=1}^{N_\mrm{B}}
\left(
\hat{\kappa}_\perp^{(b)}(T,n)
-
\overline{\kappa}_\perp(T,n)
\right)^2
\Biggr]^{1/2},
\end{aligned}
\end{equation}

with

\begin{equation}
\label{eq:kappa_bootstrap_mean}
\overline{\kappa}_\perp(T,n)
=
\frac{1}{N_\mrm{B}}
\sum_{b=1}^{N_\mrm{B}}
\hat{\kappa}_\perp^{(b)}(T,n),
\end{equation}

This is the coefficient of variations of the predictions, expressed as a percentage.\\

The region covered by the measurements shows the lowest standard deviation, indicating that the PINN produces consistent conductivity estimates when interpolating within the domain constrained with data. In contrast, the bottom-right and upper-left regions of the plot show larger deviations, since no training data are available there and the model is effectively extrapolating. The largest deviations appear at the lowest values of $n$ and $T$, where the conductivity approaches zero and its relative standard deviation therefore increases. It is also worth noting that the regions with higher $\tilde{\sigma}_\kappa$ correspond, in general, to those with larger relative errors in figure~\ref{fig:kappa_dif}. This agreement supports the use of bootstrap resampling as a practical indicator of the sampling sensitivity of the inferred conductivity.

\subsubsection{Scan in number of profiles and positions}
\label{subsubsec:scan_profiles_positions}

Experimental data acquisition in fusion devices is often costly and technically limited. It is therefore useful to estimate how the performance of the inverse PINN depends on the amount of available data, in order to identify a suitable trade-off between reconstruction accuracy and experimental effort. In the present study, the amount of training data is controlled by $N_\mrm{P}$ and $N_\mrm{x}$. The objective of this scan is therefore to determine how large $N_\mrm{P}$ and $N_\mrm{x}$ need to be to achieve a given accuracy in the inferred conductivity.\\

The performance is evaluated through the Mean Absolute Percentage Error (MAPE) of $\kappa_\perp(n,T)$, defined as
\begin{equation}
\label{eq:kappa_mae}
\mathrm{MAPE}_{\kappa}[\%]
=
\frac{1}{N}
\sum_{i=1}^{N}
\left|
\frac{\kappa_{\perp}(n_i,T_i)
-
\hat{\kappa}_{\perp}(n_i,T_i)}
{\kappa_{\perp}(n_i,T_i)}
\right|\cdot 100 ,
\end{equation}
where $\{n_i,T_i\}_{i=1}^N$ form an evenly distributed 2D grid spanning the full $(n,T)$ domain shown in figure~\ref{fig:kappa_dif}; that is, $N$ is the number of evaluation points covering the entire case space, and $\mathrm{MAPE}_\kappa$ is averaged over this whole domain rather than only the region constrained by the training data.\\

\begin{figure}
    \centering
    \includegraphics[width=1\linewidth]{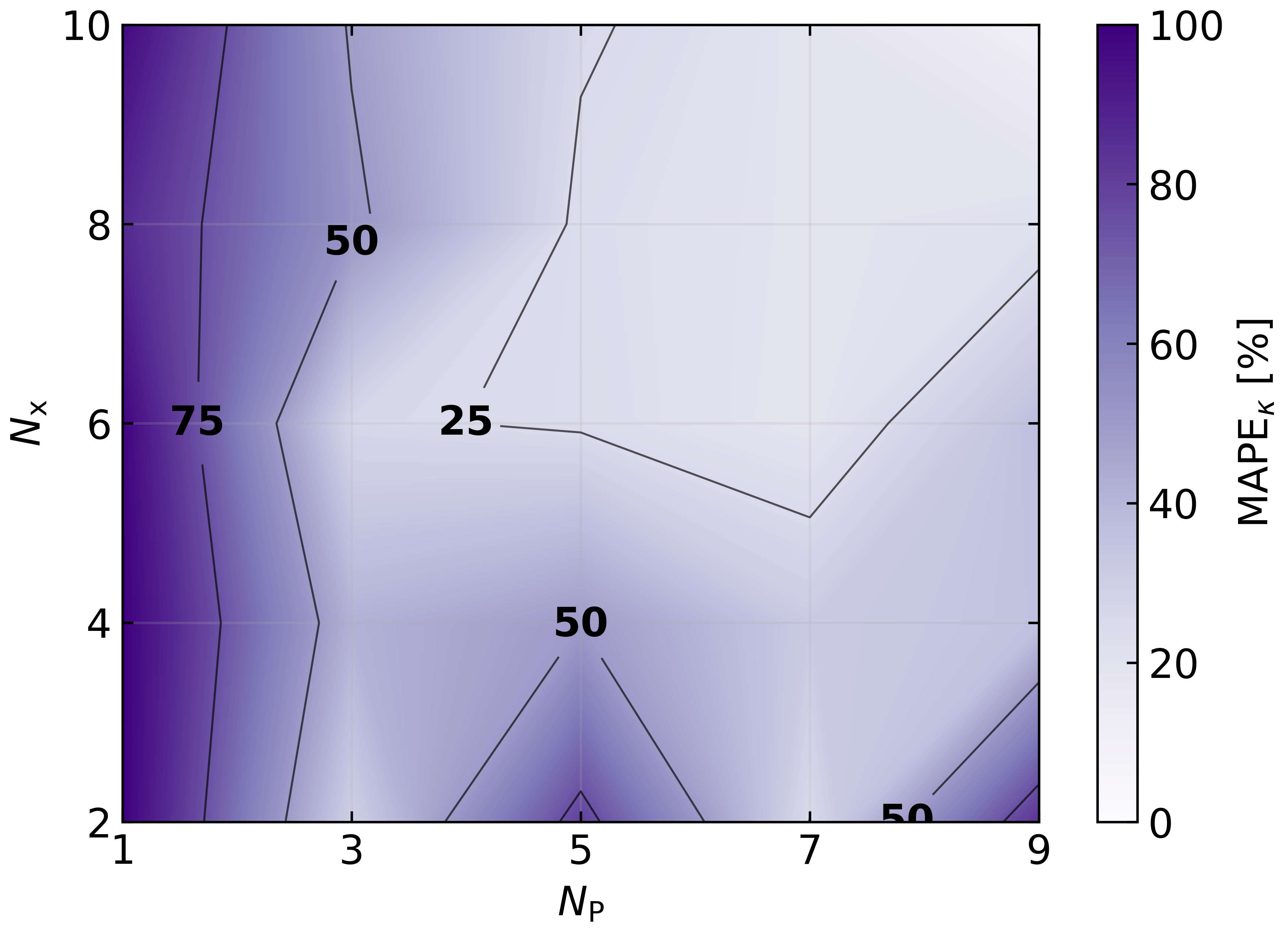}
    \caption{Mean absolute percentage error of the inferred perpendicular conductivity, $\mathrm{MAPE}_\kappa$, as a function of the number of profiles used for training, $N_\mrm{P}$, and the number of radial measurement positions per profile, $N_\mrm{x}$. For each pair $(N_\mrm{P},N_\mrm{x})$, five bootstrap realizations are performed, and the plotted value corresponds to the median value of the $\mathrm{MAPE}_\kappa$ distribution.}
    \label{fig:k_MAE_scan}
\end{figure}

Figure~\ref{fig:k_MAE_scan} shows the median of the $\mathrm{MAPE}_\kappa$ obtained for different combinations of $N_\mrm{P}$ and $N_\mrm{x}$, with five bootstrap realizations performed for each case. As expected, the largest error is found for the most weakly constrained cases, corresponding to $N_\mrm{P}=1$. Increasing both the number of profiles and the number of radial positions generally reduces the reconstruction error, although the improvement becomes less pronounced once the dataset is sufficiently informative. In this case, the error is found to saturate around $N_\mrm{P}=5$ and $N_\mrm{x}=6$, beyond which additional data provide only a marginal gain in accuracy.

\section{Application to experimental data}\label{sec:exp}

This section applies the proposed methodology to an experimental dataset. Only a limited amount of real experimental data was available at the time of this study, so the results presented here are meant to test whether the framework can be applied to real measurements, rather than to perform a complete systematic study of the parametric dependencies of $\kappa_\perp$. Applying the method to broader and better-constrained experimental datasets is left for future work. Section~\ref{subsec:setup} briefly describes the experimental setup used to obtain the measurements. Section~\ref{subsec:exp_data} presents the experimental dataset considered in the analysis. Finally, section~\ref{subsec:exp_res} shows and discusses the results obtained from the experimental application of the PINN framework.

\subsection{Experimental setup}\label{subsec:setup}

TJ-II was selected for this exploratory application because it is a well-characterized, flexible heliac-type stellarator with a long operational history \cite{alejaldre1990tj}, and because its helium-beam diagnostic provides simultaneous, spatially resolved measurements of the SOL electron density and temperature profiles \cite{branas2001atomic} required by the PINN framework. TJ-II is located at the Laboratorio Nacional de Fusión of CIEMAT, Madrid.\\

\begin{figure}
    \centering
    \includegraphics[width=1\linewidth]{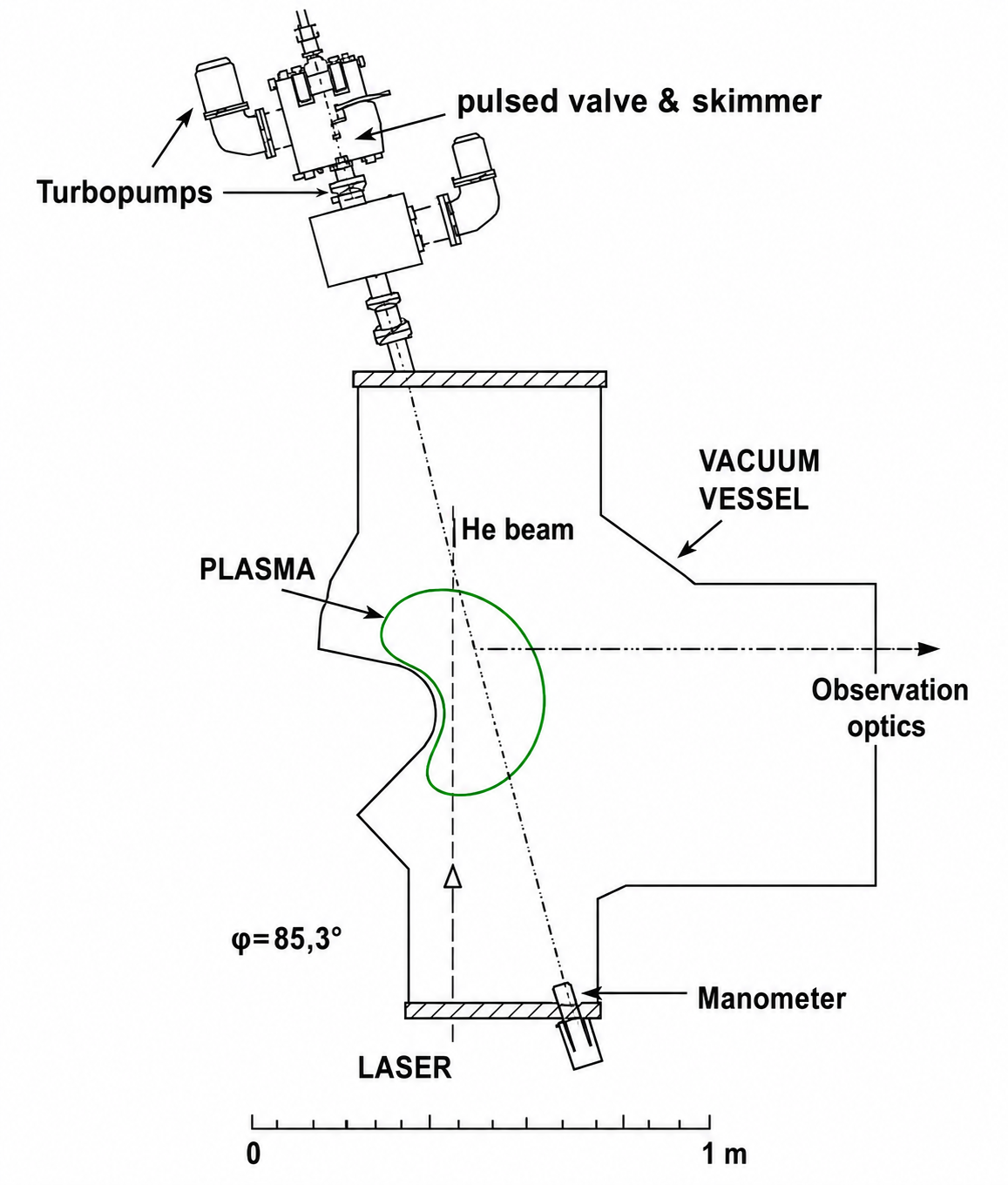}
    \caption{Schematic poloidal cross section of the helium-beam diagnostic setup. The helium-beam is injected vertically into the plasma through the vacuum vessel, while the emitted radiation is collected horizontally by the observation optics. The upper section shows the gas injection and pumping system, while the lower section indicates the manometer used to monitor the gas pressure near the injection line. Diagram adapted from \cite{branas2001atomic}.}
    \label{fig:he_beam}
\end{figure}

The helium-beam diagnostic injects a neutral helium-beam into the plasma, and the resulting helium line emission is interpreted through a collisional-radiative model to reconstruct radial profiles of the edge electron temperature and density \cite{hidalgo2005testing}. Figure~\ref{fig:he_beam} illustrates the diagnostic layout, showing the helium-beam entering the plasma from the upper part of the poloidal section. The measurement path is defined by the helium-beam trajectory, along which the neutral beam is then progressively attenuated as it penetrates into the plasma.\\

The temporal resolution of the helium-beam diagnostic is limited to $50~\mrm{ms}$, which provides $N_\mrm{R}=2$ measurements per discharge. The spatial resolution is limited to $3.5~\mrm{mm}$, corresponding to $N_\mrm{x}=4$ measurement positions within the SOL. This temporal and spatial coverage is insufficient for a complete experimental characterization of the parametric dependencies of $\kappa_\perp$ in a systematic study. Nevertheless, it supports the present exploratory objective: testing the PINN framework on real experimental profiles and obtaining an initial estimate of the inferred conductivity under realistic data constraints.

\subsection{Experimental dataset}\label{subsec:exp_data}

\begin{figure*}
    \centering
        \includegraphics[width=1\linewidth]{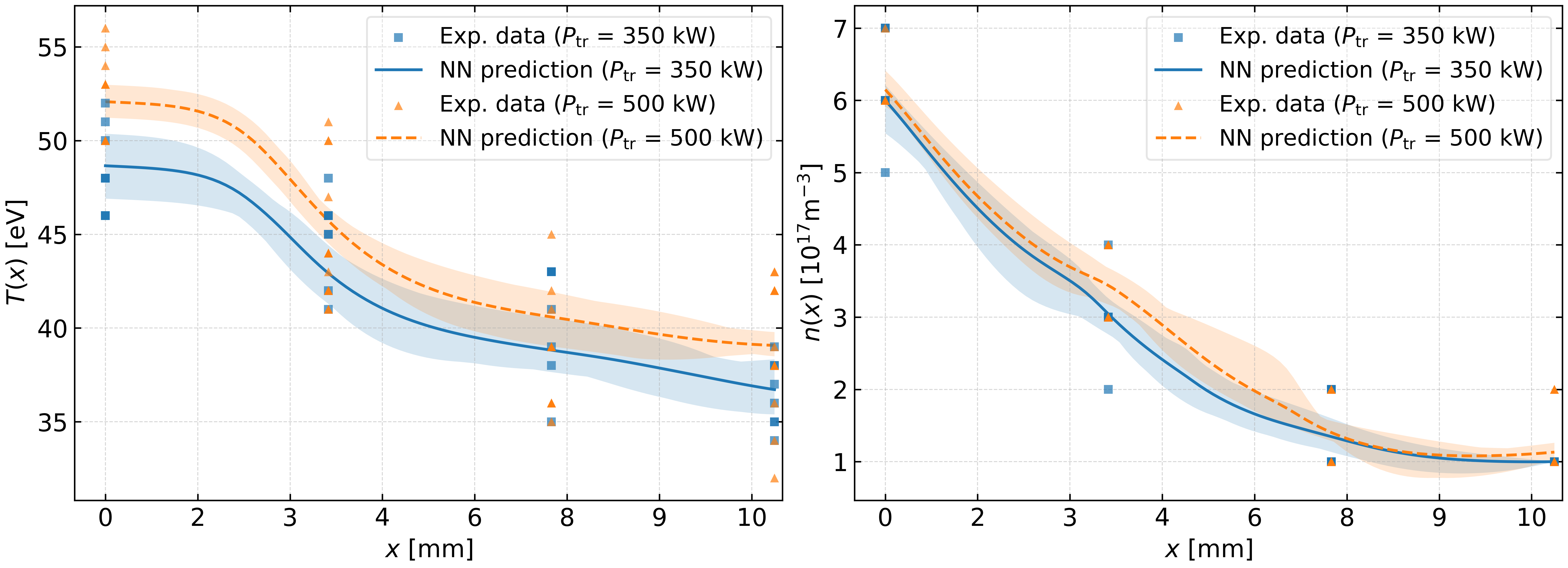}
    \caption{Comparison between the experimental dataset and the neural network reconstructions. Left: temperature profiles $T(x)$ for two transported power values. Right: density profiles $n(x)$ for the same cases. Markers correspond to the experimental measurements, with squares corresponding to $P_\mrm{tr}=350~\mrm{kW}$ and triangles to $P_\mrm{tr}=500~\mrm{kW}$. Lines represent the smooth profiles reconstructed by the neural networks averaged over all the bootstrapped runs. Shaded regions indicate the corresponding 90\% bootstrap confidence intervals, representing the uncertainty associated with variations in the sampled training data.}
    \label{fig:T_n_exp}
\end{figure*}

The experimental dataset was obtained from 13 discharges performed during the TJ-II Spring 2026 campaign. Of these, five discharges were conducted at a heating power of $P_\mrm{h}=350~\mrm{kW}$, and eight discharges at $P_\mrm{h}=500~\mrm{kW}$, using electron cyclotron resonance heating (ECRH). The line-averaged plasma density was kept at low values, $\overline{n}_e\approx 5\cdot10^{18}~\mrm{m^{-3}}$, so the measured radiation losses remained small, $P_\mrm{rad}\sim 0.01P_\mrm{h}$. Radiative losses are therefore neglected, and, under the assumed steady-state power balance, the heating power is approximated as the transported power,
\begin{equation}
P_\mrm{h}=P_\mrm{rad}+P_\mrm{tr}\approx P_\mrm{tr}.
\end{equation}

Figure~\ref{fig:T_n_exp} shows the experimental dataset for the two power values. The temperature measurements, shown with markers in the left panel, exhibit a weaker dependence on transported power than the synthetic data presented earlier. Nevertheless, higher temperatures are generally observed for the higher transported power case. The density was intended to be kept approximately constant across all discharges, and the measurements from both power cases are therefore found to be similar, as reflected in the right panel.\\

The localization of the LCFS typically has an uncertainty of $5$--$10~\mrm{mm}$ \cite{lopez2016monte}. Therefore, in the present analysis, the LCFS position is estimated using a heuristic profile-based criterion, in which the LCFS is identified as the first point among the last four measurements for which both $T$ and $n$ show a clear decreasing trend. This criterion provides a simple and consistent way to select reasonable SOL data to feed the PINN, adequate for the exploratory purposes of this work.\\

It was necessary to estimate the perpendicular conductivities at the LCFS for the two different powers in order to calculate the loss function, equation~\ref{eq:total_loss_pinn}. To obtain these values, the experimental temperature profiles were pre-fitted in the vicinity of the LCFS. The derivative $d T/d x$ at $x=0$ was then computed from the fitted profiles and inserted into equation~\ref{eq:LCFS_heat_flux}, allowing $\kappa_\perp^\mrm{LCFS}$ to be solved directly. Using this procedure, the LCFS conductivity values obtained to constrain the PINN were
$\kappa_\perp^\mrm{LCFS}=(12.5\pm 0.8)\cdot10^{19}~\mrm{m^{-1}s^{-1}}$
for $P_\mrm{tr}=350~\mrm{kW}$ and
$\kappa_\perp^\mrm{LCFS}=(16.1\pm 0.7)\cdot10^{19}~\mrm{m^{-1}s^{-1}}$
for $P_\mrm{tr}=500~\mrm{kW}$. The reported uncertainties originate from the standard errors of the fitted profiles and are propagated to $\kappa_\perp^\mrm{LCFS}$ through equation \ref{eq:LCFS_heat_flux}.\\

\subsection{Experimental results}
\label{subsec:exp_res}

\begin{figure*} \centering \includegraphics[width=1\linewidth]{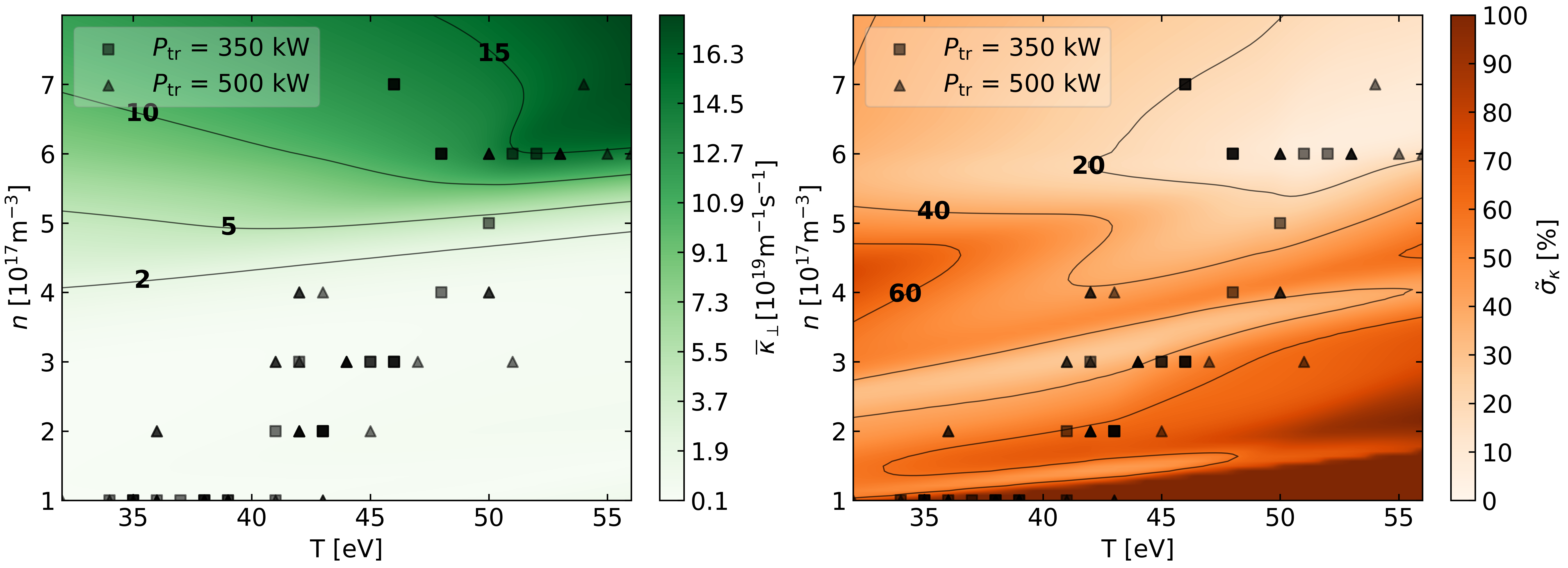} \caption{Inferred perpendicular conductivity from the TJ-II experimental dataset. Left: ensemble mean conductivity, $\overline{\kappa}_\perp(T,n)$, obtained from the bootstrap realizations. Right: normalized ensemble standard deviation, $\tilde{\sigma}_\kappa(T,n)$, expressed as a percentage. Markers indicate the experimental data points used for training, with squares corresponding to $P_\mrm{tr}=350~\mrm{kW}$ and triangles to $P_\mrm{tr}=500~\mrm{kW}$.} \label{fig:kappa_exp} \end{figure*}

The PINN framework was applied to the experimental dataset described in section~\ref{subsec:exp_data}, using the same reduced transport model and hyperparameters validated with synthetic data. Since the ground truth conductivity is not known for experimental measurements, the results are interpreted through the consistency of the reconstructed profiles and the variability of the inferred conductivity across bootstrap realizations.\\

The PINN model was trained and evaluated for $N_\mrm{B}=10$ independent bootstrap realizations. As done in section~\ref{sec:synthetic_validation}, the experimental dataset was resampled for each bootstrap realization by drawing $(T,n,x,P_\mrm{tr})$ tuples with replacement from the pooled dataset. In addition, the LCFS conductivity associated with each transported power value was independently sampled from a Gaussian distribution centered on its estimated value, with a standard deviation equal to the corresponding reported uncertainty. This additional sampling propagates the uncertainty in the LCFS conductivity constraints through the bootstrap analysis.\\

Figure~\ref{fig:T_n_exp} shows, with solid and dashed lines, the bootstrap-averaged PINN reconstructions $\hat{T}(x)$ and $\hat{n}(x)$, while shadow regions represent the $90\%$ confidence intervals for the two transported power values. The reconstructed profiles reproduce the experimental data reasonably well, and the confidence intervals show reproducible computations, with variations of $\pm2~\mrm{eV}$ in temperature and $\pm0.5~\mrm{10^{17}m^{-3}}$ in density. For $x \gtrsim 8~\mrm{mm}$, density profiles become weakly varying and they reach low values, providing a first estimate of the radial extent of the measured SOL region.\\

The physics-consistency score $\mathcal{S}_f$ applied to the bootstrap ensemble yields $\mathcal{S}_f=0.814$. This value is considerably lower than that obtained in the synthetic case, where $\mathcal{S}_f=0.998$. This difference may indicate that the transport physics underlying the experimental data are not fully captured by equation~\ref{eq:1D_eq_pinn}, in contrast to the synthetic case, where the governing equation is satisfied by construction. Given the limited experimental dataset, this lower score should be interpreted as an exploratory measure rather than as a calibrated acceptance criterion for the reduced transport model.\\

Figure~\ref{fig:kappa_exp} shows the ensemble mean of the inferred conductivity, $\overline{\kappa}_\perp(T,n)$, obtained from the 10 bootstrap realizations, together with its normalized ensemble standard deviation, $\tilde{\sigma}_\kappa(T,n)$, over the evaluated $(T,n)$ domain. As in the synthetic validation, the inferred conductivity should be interpreted primarily within the region of the $(T,n)$ space covered by the measurements, where the PINN is constrained by both the labelled data and the transport equation. The largest deviations are indeed found in the lower-right corner of the plot, where experimental data are absent.\\

Focusing on the left panel of figure~\ref{fig:kappa_exp}, the inferred conductivity ranges approximately between $10^{18}$ and $10^{20}~\mrm{m^{-1}s^{-1}}$. This corresponds to a gyro-Bohm-normalized conductivity of approximately $\kappa_\perp/\kappa_\perp^\mrm{gB}\sim800$. For comparison, values around $\kappa_\perp/\kappa_\perp^\mrm{gB}\sim100$ have been reported for the Wendelstein 7-X (W7-X) stellarator \cite{bold2024impact,killer2021turbulent}. The gyro-Bohm conductivity used for this normalization is computed as
\begin{equation}
    \kappa_\perp^\mrm{gB}
    =
    \frac{nv_\mrm{th,i}^3}{\Omega^2_\mrm{i}a}
    =
    \sqrt{\frac{8m_\mrm{i}}{e^4}}
    \frac{nT^{3/2}}{B^2a},
\end{equation}
where $v_\mrm{th,i}=\sqrt{2T_\mrm{i}/m_\mrm{i}}$ is the ion thermal velocity and $\Omega_\mrm{i}=eB/m_\mrm{i}$ is the ion gyrofrequency. For the TJ-II gyro-Bohm reference value, $B=1~\mrm{T}$ and $a=0.22~\mrm{m}$ are adopted, together with a representative SOL temperature $T=40~\mrm{eV}$. Subject to compatible definitions and normalizations, the normalized conductivity inferred for TJ-II is roughly eight times larger than the values reported for W7-X. This difference may partly reflect variations in the underlying SOL transport regimes, highlighting the need for a broader multi-machine analysis to better characterize and understand perpendicular heat transport across different stellarator devices.\\

Regarding the dependencies of $\overline{\kappa}_\perp(T,n)$, the maximum values are found at high $T$ and high $n$, corresponding approximately to the LCFS region. From this region outward, the inferred conductivity decreases as both $T$ and $n$ decrease, in qualitative agreement with the trends expected from Bohm-like or gyro-Bohm-like scalings. However, for $n<4\cdot 10^{17}~\mrm{m^{-3}}$, the temperature dependence appears to weaken. This behaviour is not directly explained by the reference scalings discussed in section~\ref{subsubsec:kperp_scalings}. It may instead reflect the increased relative uncertainty of the inferred conductivity, or the fact that $\overline{\kappa}_\perp$ represents an effective transport coefficient that absorbs physics not explicitly included in the reduced 1D model. As stated above, these results should not be interpreted as a definitive analysis of $\kappa_\perp$ and its parametric dependencies, but rather as a first exploratory application of the method to real experimental data. A broader and better-constrained experimental dataset will be required to validate these trends and perform a complete and systematic transport analysis.\\

\section{Conclusions}
\label{sec:conclusions}

This work has developed and tested an inverse PINN framework to infer the perpendicular heat conductivity, $\kappa_\perp(n,T)$, in the SOL of magnetically confined fusion plasmas. The model combines sparse measurements of temperature and density profiles with a reduced 1D transport equation, allowing the conductivity to be inferred as a functional dependence on the local plasma density and temperature. The proposed architecture uses three neural networks: two reconstruct $\hat{T}(x,P_\mrm{tr})$ and $\hat{n}(x,P_\mrm{tr})$, while a third represents $\hat{\kappa}_\perp(T,n)$. By including the transported power as an input, the model can be trained simultaneously on several plasma profiles obtained under different heating conditions.\\

The method was first validated using synthetic data generated from a prescribed conductivity function. In this controlled case, the PINN recovered the imposed dependence of $\kappa_\perp(n,T)$, with errors below $10\%$ in the data-constrained region and a differential equation satisfaction metric of $\mathcal{S}_f=0.998$. Bootstrap resampling provided a useful indicator of prediction consistency and reliability, with larger variability appearing in weakly sampled regions of the $(T,n)$ domain. A scan in the number of profiles and radial measurement positions showed that the reconstruction accuracy improves with data availability, with the error saturating around $N_\mrm{P}=5$ and $N_\mrm{x}=6$ for the synthetic conditions considered.\\

The framework was then applied to a TJ-II experimental dataset obtained with the helium-beam diagnostic. This application was intended as a first exploratory test with real data, rather than a complete systematic study of $\kappa_\perp$ dependencies. The reconstructed profiles reproduced the experimental measurements reasonably well, with a physics-consistency score of $\mathcal{S}_f=0.82$. The inferred conductivity was found in the range $10^{18}$--$10^{20}~\mrm{m^{-1}s^{-1}}$, corresponding to a gyro-Bohm-normalized value of order $\kappa_\perp/\kappa_\perp^\mrm{gB}\sim800$. This value is roughly eight times larger than that reported for W7-X, a contrast that possibly reflects differences in the underlying SOL transport regimes between the two devices, and which motivates a broader multi-device analysis.\\

Overall, the results indicate that inverse PINNs can provide a useful framework for estimating effective SOL transport coefficients from sparse experimental profiles. However, the inferred $\kappa_\perp(n,T)$ should be interpreted within the assumptions of the reduced 1D model and mainly in the region of the $(T,n)$ domain covered by data. In this work, the uncertainty of the inferred conductivity was quantified through bootstrap resampling of the training data (section~\ref{subsec:numerical_results} and figure~\ref{fig:kappa_exp}). A more thorough treatment of this uncertainty could be obtained within a Bayesian inference formalism, which would in addition provide a natural tool for experimental design: given an approximate prior estimate of $\kappa_\perp(n,T)$, either from a previous inference or from a limited set of experimental values, the resulting posterior uncertainty map -- analogous to that shown in figure~\ref{fig:kappa_exp} -- could be used to identify the $(T,n)$ regions where new measurements would most effectively reduce the uncertainty, thereby helping to inform the design of future diagnostic campaigns. Future work should therefore explore this Bayesian extension, apply the method to broader experimental datasets, and refine the treatment of geometry and parallel losses to extend its applicability to larger devices such as W7-X and LHD.

\section{Acknowledgements}
This work was partially funded by the MINERIA project (Grant No.~PID2024-157169OB-I00) and the INEXTELA project (Grant No.~PID2025-169524OB-I00). Computational resources were provided by the Extremadura Research Centre for Advanced Technologies (CETA-CIEMAT), funded by the European Regional Development Fund (ERDF). CETA-CIEMAT belongs to CIEMAT and the Government of Spain. The author gratefully acknowledges the Harvard John A. Paulson School of Engineering and Applied Sciences for hosting a two-month research stay that contributed to the development of this work.


\printbibliography 


\end{document}